\documentclass[doublecol]{epl2}

\usepackage{amssymb,amsmath,latexsym}

\usepackage{epsfig,psfrag,subfigure}
\usepackage{graphicx,psfrag,subfigure}
\usepackage{color}
\usepackage{bbold}

\newcommand\bea{\begin{eqnarray}}
\newcommand\eea{\end{eqnarray}}
\newcommand\beq{\begin{equation}}
\newcommand\eeq{\end{equation}}
\newcommand\bsq{\begin{subequations}}
\newcommand\esq{\end{subequations}}

\newcommand{\noi}{\noindent}

\def\nn{\nonumber}
\def\f{\frac}

\def\om{\omega}
\def\bom{{\bar \omega}}
\def\Om{\Omega}
\def\de{\delta}
\def\ep{\epsilon}

\def\si{\sigma}
\def\Do{\partial}

\def\vr{\varrho}

\def\la{\langle}
\def\ra{\rangle}
\def\mbb{\mathbb}

\title{\textbf{Conductance of Tomonaga-Luttinger liquid wires and junctions 
with resistances}}
\author{Abhiram Soori \and Diptiman Sen}
\institute{Centre for High Energy Physics, Indian Institute of Science, 
Bangalore 560012, India}

\date{\today}

\abstract
{We study the effect that resistive regions have on the conductance of a 
quantum wire with interacting electrons which is connected to Fermi liquid 
leads. Using the bosonization formalism and a Rayleigh dissipation function 
to model the power dissipation, we use both scattering theory and
Green's function techniques to derive 
the DC conductance. The resistive regions are generally found to lead to 
incoherent transport. For a single wire, we find that the resistance adds in 
series to the contact resistance of $h/e^2$ for spinless electrons, and the 
total resistance is independent of the Luttinger parameter $K_W$ of the wire. 
We numerically solve the bosonic equations to illustrate what happens when a 
charge density pulse is incident on the wire; the results depend on the 
parameters of the resistive and interacting regions in interesting ways. 
For a junction of Tomonaga-Luttinger liquid wires, we use a dissipationless 
current splitting matrix to model the junction. For a junction of three wires 
connected to Fermi liquid leads, there are two families of such matrices; 
we find that the conductance matrix generally depends on $K_W$ for one 
family but is independent of $K_W$ for the other family, regardless of the
resistances present in the system.}


\begin{document}

\maketitle

{\it Introduction.-}
It is well known that for non-interacting electrons, the conductance of a 
narrow ballistic quantum wire is quantized in units of $e^2/h$ at low
temperatures~\cite{buttiker,wees}. This remains true when electron-electron 
interactions are taken into account in the wire, provided that there are no 
sources of backscattering (such as impurities) and that the wire is 
connected to leads where there are no interactions 
\cite{maslov,ponomarenko,safi1,safi3,thomale}. Namely, if 
the wire is modeled as a Tomonaga-Luttinger liquid (TLL) and the interaction 
strength is given by the Luttinger parameter $K_W$, the conductance of a clean
wire is independent of $K_W$. This breaks down if there are isolated 
impurities in a wire with interacting electrons; the impurity strengths 
then satisfy some renormalization group (RG) equations, and the conductance
depends on $K_W$ and other parameters like the wire length, the distances 
between the impurities, and the temperature~\cite{kane,furusaki,expt}.
One can think of the impurities as giving rise to a resistance which leads
to power dissipation, although this aspect is usually not highlighted in the
literature. There have been some studies of power dissipation on the edges 
of a quantum Hall system~\cite{wen} and also at a junction of quantum wires 
due to the presence of bound states~\cite{bellazzini2}. However, there has 
been relatively little discussion of the effects of an extended region of 
dissipation (a patch of resistance) within the framework of TLL theory or 
bosonization which is well-suited for studying the effects of interactions 
between electrons~\cite{boson}. Such a theory would have the benefit of 
combining the wealth of knowledge of TLLs with the classical notion of 
resistance. Further, a large amount of work has been done on junctions of 
several quantum wires theoretically 
\cite{sandler,nayak,lal,chen,chamon,meden,das,giuliano,bellazzini1,agarwal}
and experimentally~\cite{fuhrer,terrones}, and it would be useful to know what
effect resistances in the wires have on the conductance matrix of such a 
system. A junction of three quantum wires with interacting spin-1/2 electrons 
has been studied in Ref.~\cite{hou}, and it has been found that some of the 
fixed points of the RG equations have different properties for the charge 
and spin sectors. In this context, we would like to mention the work in 
Ref.~\cite{rech}. Here the effect of an extended region of inhomogeneity in 
a quantum wire has been studied, and it has been shown that this leads to
weak backscattering which gives rise to a resistance which is linear in
the temperature. Further, the resistances for the charge and spin sectors
are different; the sum of the two gives the total resistance.

In this paper, we will use the technique of bosonization to study the 
effect of patches of resistance on the conductance of a quantum wire system 
with or without junctions. Our treatment will be classical in the sense that
the resistance will be taken to be purely a source of Ohmic power dissipation;
we will not consider the microscopic origins of the resistance such as
point impurities which can scatter the electrons quantum mechanically. As 
a result, the transport will be seen to be incoherent, with the resistance 
of different patches adding in series with no effects of interference;
the incoherence also implies that RG equations will
play no role in the analysis. Using the idea of a Rayleigh dissipation 
function~\cite{goldstein} to model the resistance patches, we will obtain the
equations of motion for the bosonic field whose space and time derivatives 
give the electron charge density and current respectively. We then use both 
a scattering solution of the equations of motion~\cite{safi1,safi3} and
a Green's function approach~\cite{maslov} to obtain the 
DC conductance $G$. Our analysis leads to several new results which are as 
follows. For a single wire, we calculate $G$ when the Luttinger parameter 
$K$, the velocity $v$ and the resistivity $r$ all vary with the spatial 
coordinate $x$ in the region of the quantum wire. 
The equations of motion enable us to numerically study 
the space-time evolution when a charge density wave of arbitrary shape is 
incident on the wire region. For the case of several quantum wires meeting 
at a junction, we model the junction using an orthogonal and
dissipationless current splitting matrix $M$. For a three-wire junction,
it is known that there are two families of $M$ which have determinant $\pm 1$
respectively. When resistance patches are then introduced in each of the wires
some distance away from the junction, the conductance matrix $G$ of the system
is generally found to depend on the matrix $M$ as well as some of the 
parameters mentioned above; this will be discussed in detail. For simplicity,
we will restrict our analysis to the case of spinless electrons.

{\it Equation of motion.-}
We begin by studying a single wire with interacting spinless electrons.
In the absence of backscattering processes, the bosonic Lagrangian is given by
\beq L ~=~ \int_{-\infty}^\infty dx ~[\frac{1}{2vK} (\Do_t \phi)^2 ~-~
\frac{v}{2K} (\Do_x \phi)^2], \label{lag} \eeq
where $K$ and $v$ denote the Luttinger parameter and velocity respectively;
these parameters can vary with $x$ within a finite region which we will take 
to be $-L/2 < x < L/2$. The Fermi liquid leads will be assumed to lie in the 
regions $|x| > L/2$, where $v=v_F$ and $K=1$ are constant; in the leads,
the frequency and wave number of a plane wave are related as $\om = v_F |k|$. 
The electron charge density $n$ and current $j$ are given in terms of 
the bosonic field as $n = -e \Do_x \phi/\sqrt{\pi}$ and $j = e \Do_t \phi/
\sqrt{\pi}$, where $e$ is the electron charge; these densities clearly satisfy
the equation of continuity $\Do_t n + \Do_x j = 0$. The energy of the system 
is then given by 
\beq E ~=~ \int_{-\infty}^\infty dx ~[\frac{1}{2vK} (\Do_t \phi)^2 ~+~
\frac{v}{2K} (\Do_x \phi)^2]. \label{energy} \eeq
We now introduce dissipation in the model through a Rayleigh dissipation 
function
\beq {\cal F} ~=~ \frac{1}{2} ~\int_{-\infty}^\infty dx ~r ~j^2, 
\label{diss} \eeq
where the resistivity $r$ can also vary with $x$ but will be taken to be 
non-zero only within the region $|x| < L/2$. The function in eq.~\eqref{diss}
contributes to the equation of motion as $d/dt (\de L/\de \Do_t \phi) - \de L/
\de \phi + \de {\cal F}/\de \Do_t \phi = 0$~\cite{goldstein}, which gives
\beq \frac{1}{vK} ~\Do_t^2 \phi ~-~ \Do_x (\frac{v}{K} \Do_x \phi) ~+~
\frac{e^2}{\pi} ~r ~\Do_t \phi = 0. \label{eom} \eeq
(Note that we have set $\hbar =1$, so that $e^2/(2\pi) = e^2/h$).
One can then show that the power dissipation is given by 
\beq \frac{dE}{dt} ~=~ - \int_{-\infty}^\infty dx ~\Do_t \phi ~\de {\cal F}/
\de \Do_t \phi \eeq
which is equal to $- j^2 R$ in a steady state as desired; here $R = 
\int_{-\infty}^\infty dx ~r$ is the total resistance, and steady state means 
that $j$ is independent of $x$ (this follows from the equation of 
continuity and the fact that $\Do_t n =0$ in a steady state).

One can compute the conductance of the system in two ways. The first way is 
to consider the interacting and resistive regions as sources of scattering,
as has been done for an interacting region in Refs.~\cite{safi1,safi3}.
We allow a plane wave with frequency $\om$ to be incident on this
region from the left, and compute the reflection and transmission amplitudes 
as functions of $\om$. The latter amplitude is related, in the limit $\om \to 
0$, to the dc conductance $\si_{dc}$. The second way is to compute the 
Fourier transform of the Green's function in imaginary time and hence the 
nonlocal ac conductance; this again gives $\si_{dc}$ in the limit $\om \to 0$
\cite{maslov}. We will use both these methods, the scattering method for
a single wire and the Green's function for a junction of three wires.

{\it Transmission through a dissipative region.-}
To illustrate the scattering method for computing $\si_{dc}$
\cite{safi1,safi3}, let us first
consider a non-interacting system in which $K=1$ and $v=v_F$ are independent
of $x$, while $r(x) = r_0$ for $-a<x<a$ and $0$ elsewhere. This describes a 
dissipative region $(-a,a)$ connected to leads on the two sides. For a 
plane wave incident from the left with $k=\om /v_F$, the spatial part of 
the solution $\phi_k (x,t) = f_k (x) e^{-i\om t}$ is given by
\bsq \label{sktk} \begin{align}
f_k &=& e^{ikx} + s_ke^{-ikx} ~~~~~~~~~~~~~~~~~{\rm for}~~x\le-a, \label{sktka} \\
&=& t'_{k'}e^{ik'x} + s'_{k'}e^{-ik'x} ~~~~{\rm for}~~-a \le x\le a, \label{sktkb} \\
&=& t_ke^{ikx}~~~~~~~~~~~~~~~~~~~~~~~~~~~~~~{\rm for}~~a\le x. \label{sktkc}\end{align}
\esq
Using eq.~\eqref{eom}, and the continuity of $f_k$ and $\Do_x f_k$ at $x= 
\pm a$, we find that $s_k$ and $t_k$ are given by
\bea s_k &=& \f{(\eta^2-1)(e^{i2\eta ka}-e^{-i2\eta ka})}{(1+\eta)^2-(1-\eta)^2
e^{i4\eta ka}}, \nn \\
t_k &=& \f{4\eta e^{i2 ka(\eta-1)}}{(1+\eta)^2-(1-\eta)^2e^{i4\eta ka}},
\label{tk} \\
{\rm where}~~~ \eta &=& \f{k'}{k} = \sqrt{1+\f{\zeta}{k}}, ~~{\rm and}~~ 
\zeta = i\f{e^2}{\pi} r_0. \label{eta} \eea

We now consider what happens when a $\de$-function charge density
pulse is incident on the dissipative region from the left lead. 
At $t=0$, the pulse is at $x_0 < -a $ with velocity $+v_F$. 
This pulse is described by: 
\beq \phi (x,t) ~=~ i \sqrt{\pi} ~\int_{-\infty}^\infty ~\frac{dk}{2\pi} ~
\frac{e^{ik(x-x_0-v_Ft)}}{k+i\ep}, \eeq
so that $n = e \de (x-x_0-v_F t)$ and $j = ev_F \de (x-x_0-v_F t)$ for $x<-a$ 
and $t < -(a+x_0)/v_F$ (before scattering into the resistive region). 
For $x>a$, the corresponding current is given by
\beq j(x,t) ~=~ ev_F ~\int_{-\infty}^\infty ~\frac{dk}{2\pi} ~t_k ~
e^{ik(x-x_0-v_Ft)}. \eeq
The nonlocal ac conductivity is given by $\si(x,x_0,t) = e j(x,t)/(2\pi)$, 
where $x>a$ and $x_0 < -a$. The DC conductance is given by the zero 
frequency limit of the Fourier transform,
\bea \si_{dc} &=& 
\lim_{\Om \to 0^+} \int_{-\infty}^{\infty}dt ~e^{i\Om t} \si (x,x_0,t) \nn \\
&=& \f{e^2}{2\pi} ~\lim_{k\to 0^+} t_k \nn \\
&=& \f{e^2}{2\pi} ~\lim_{k\to 0^+} \Big[\f{4\eta}{(1+\eta)^2} ~
\f{e^{i2(\eta-1) ka}}{1- \chi e^{i4\eta ka}}\Big], \eea
where $\chi = (1-\eta)^2 /(1+\eta)^2$.
Using $\lim_{k \to 0} \eta = \sqrt{\zeta/k}$ and $\lim_{k \to 0} 
\chi = 1-4\sqrt{k/\zeta}$, we find that 
\beq \si_{dc} ~=~ \f{e^2}{2\pi} ~\f{1}{1-i\zeta a} ~=~\f{e^2}{2\pi} ~
\f{1}{1 +\f{e^2R}{2\pi}}, \label{dc-cond} \eeq
where $R=2ar_0$ is the total resistance. This expression shows that $R$ adds
in series to the contact resistance of $2\pi/e^2$; this is the expected 
property of resistance in a phase incoherent system. We obtain the same 
result for $\si_{dc}$ through the Green's function method outlined below 
for a three-wire system. 

The expression in eq.~\eqref{dc-cond} can be derived for a general 
resistance profile $r(x)$ as follows. For $\om = 0$, a solution of
eq.~\eqref{eom} is $\phi = c$, where $c$ is a constant. Let us now 
look for a solution  which is valid upto first order in $\om$ 
and $k=\om/v_F$, and which reduces to $\phi = c$ in the limit $\om \to 0$. 
Assuming that $\phi_k (x,t) = f_k (x) e^{-i\om t}$, where $f_k$ has the
forms given by eq.~\eqref{sktka} and eq.~\eqref{sktkc} for $x < -a$
and $>a$ respectively, we must have $1+s_k = t_k = c$ to zero-th order 
in $\om$ as $\om \to 0$. Next, on ignoring the term of order $\om^2$ 
in eq.~\eqref{eom}, we obtain 
\beq -~ \Do_x (\frac{v}{K} \Do_x f_k) ~-~ i \om \frac{e^2}{\pi} ~r ~f_k ~=~ 0.
\label{eom2} \eeq
Replacing $f_k$ by the constant $c = t_k$ in the second term in 
eq.~\eqref{eom2} (we can do this since that term has a factor of $\om$), 
and integrating that equation from $x=-a-\ep$ to $a+\ep$, we obtain 
$ik v_F [t_k - (1-s_k)] = - i \om t_k (e^2 R/\pi)$, where $R = \int_{-a}^a 
dx ~r (x)$ is the total resistance. Combining this with $1+s_k = t_k$, 
we obtain
\beq t_{k\to 0} ~=~ \f{1}{1 +\f{e^2R}{2\pi}}, \eeq
from which the result for $\si_{dc} = (e^2/2\pi) t_{k \to 0}$ follows.

It is interesting to compare the evolution of a charge density pulse incident 
on a dissipative region in a non-interacting system with the evolution of the 
same pulse in a non-dissipative but interacting system ($K_W \ne 1$). [The 
latter case was studied in Refs.~\cite{safi1,safi3}. It was shown there 
that a series of pulses emerges on both sides of the wire, such that eventually
the integrated pulse on the left (i.e., the total reflection probability) is 
zero, while the integrated pulse on the right (i.e., the total transmission 
probability) is unity]. We have time evolved eq.~\eqref{eom} numerically; a 
von Neumann stability analysis was performed to ensure that the numerical 
errors remain small. 

\begin{figure}[htb]
\begin{center} \epsfig{figure=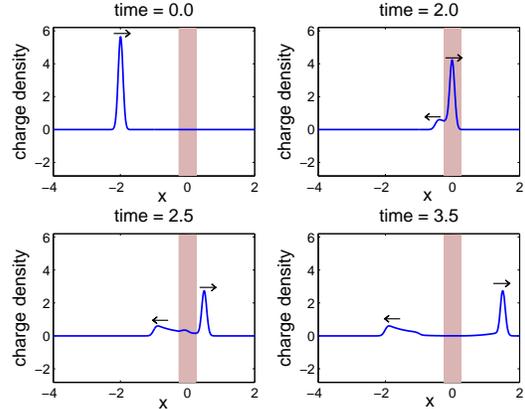,width=8.0cm}
\end{center}
\caption{Evolution of a charge density pulse incident on a dissipative region
(shaded), with $l_0 = 0.1$ (the width of the pulse), $a=0.25$ and $r_0 = 3$.} 
\label{evol_Resist} \end{figure}
\begin{figure}[htb]
\begin{center} \epsfig{figure=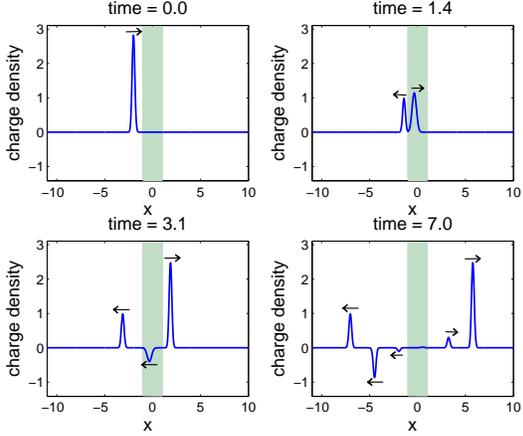,width=8.0cm} \end{center}
\caption{Evolution of a charge density pulse incident on an interacting region
(shaded), with $l_0 = 0.2$ (the width of the pulse), $a=1$, $K_W = 0.6$ and
$v_W = 1.6$.} \label{evol_LL} \end{figure}
Figures \ref{evol_Resist} and \ref{evol_LL} show the 
density profiles at different times for the purely dissipative and purely
interacting cases respectively; in the first case, we have chosen $K=1$ and 
$v=1$ everywhere, while in the second case, we have chosen $K=1$ and $v=1$ 
in the leads, but $K=0.6$ and $v=1.6$ in the interacting region. In
fig.~\ref{evol_Resist}, one sees only one reflected and one transmitted pulse. 
The width of the reflected pulse ($=4a$) is equal to twice the length of 
the dissipative region, providing us with an insight into the nature of 
dissipation. Namely, a pulse gets reflected from each point in a dissipative 
region; this makes the width of the reflected pulse (i.e., the distance 
between waves being reflected from the left and right ends of that region) 
equal to $4a$. On the other hand, when a pulse approaches an interacting 
region in which $K$ is piecewise constant, it gets reflected only from the 
points of inhomogeneity, i.e., where $dK/dx$ is not zero. Fig.~\ref{evol_LL}
shows a series of reflected and transmitted pulses in agreement with the 
results obtained analytically in Refs.~\cite{safi1,safi3}.

\begin{figure}[htb]
\begin{center} \epsfig{figure=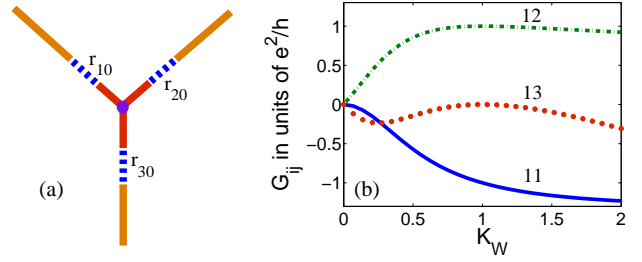,width=8.7cm} \end{center}
\caption{(a) A schematic diagram of a three-wire junction with an interacting 
region close to the junction (red solid region), a dissipative region (blue 
dotted region further from the junction), and Fermi liquid leads (brown solid 
region furthest from the junction). (b) The matrix elements $G_{ij}$, with 
$ij=$ 11, 12 and 13, are shown as functions of $K_W$ for the case of $M_1$ 
with $\theta = 2\pi/3$ and $r_{i0} = 0$ for $i=1,2,3$.} \label{diagram} 
\end{figure}

{\it Green's function calculation for three-wire junction.-}
We now consider a junction of three dissipative TLL wires as shown 
in fig.~\ref{diagram} (a), each of which contains three regions: 

\noi (i) $0\le x_i \le L_{i1} \ne 0 $ --- the region around the junction where
$K(x_i)=K_W$; elsewhere $K(x_i)= 1$, 

\noi (ii) $L_{i1} \le x_i \le L_{i2}$ --- a dissipative region where $r (x_i)=
r_{i0}$; elsewhere $r (x_i)=0$, and

\noi (iii) $x_i \ge L_{i2}$ --- semi-infinite leads.

Here $i$ labels the wires, and on wire $i$, the coordinate $x_i$ runs from 
$0$ to $\infty$, with $x_i =0$ corresponding to the junction point. The
regions $x_i \ge L_{i2}$ model the two- or three-dimensional leads which are 
assumed to be Fermi liquids with no interactions between the electrons; hence
we set $K=1$ in those regions.

Following Ref.~\cite{maslov}, we write 
\bea I_i &=& \sum_{j=1}^{3} \int_0^{L_{j2}} dx_j' \int_{-\infty}^\infty 
\f{d\om}{2\pi} e^{-i\om t} \si_{ij,\om} (x_i,x_j') E_{\om}(x_j'), \nn \\
\label{kubo} \eea
in the linear response regime, where $E_{\om}(x_j')$ is the Fourier component 
of the electric field $E(x_j',t)$ on wire $j$, and $\si_{ij,\om} (x_i,x_j')$ 
is the nonlocal conductance matrix. We then obtain 
\beq \si_{ij,\om} (x_i,x_j') ~=~ -\f{e^2 \bom}{\pi} ~{\cal G}_{ij,\bom}(x_i,
x_j'), \label{sigma-G} \eeq
where $\bom = - i \om$, and
\beq {\cal G}_{ij,\bom} (x_i,x_j') ~=~ \int_{-\infty}^\infty \f{d\tau}{
2\pi} \la T^*_{\tau} \phi_i (x_i,\tau) \phi_j (x_j',0) \ra e^{-i\bom \tau} 
\label{G} \eeq
is the propagator of the bosonic field in imaginary time, $\tau = it$.
The Green's function satisfies the equation
\bea & & \Big[ -\Do_{x_i} \Big(\f{v(x_i)}{K(x_i)} \Do_{x_i} \Big) + 
\f{\bom^2}{v(x_i)K(x_i)} - \f{e^2 \bom}{\pi} r (x_i) \Big] \nn \\
& & ~~{\cal G}_{ij,\bom} (x_i,x_j') ~=~ \de_{ij} ~\de (x_i-x_j'), 
\label{G-Eq-Mo} \eea 
with the following boundary conditions:

\noi (i) ${\cal G}_{ij,\bom} (x_i,x_j')$ is continuous at $x_i=x_j'$ (where 
$0<x_j'< L_{j2}$) and $-\f{v(x_i)}{K(x_i)} \Do_{x_i} {\cal G}_{ij,\bom} 
(x_i,x_j')|_{x_j'-\ep}^{x_j'+\ep} = \de_{ij}$,

\noi (ii) ${\cal G}_{ij,\bom}(x_i,x_j')$ and $- \f{v(x_i)}{K(x_i)} \Do_{x_i} 
{\cal G}_{ij, \bom}(x_i,x_j')$ are continuous at $x_i=L_{i1}~{\rm and}~L_{i2}$,

\noi (iii) if ${\cal G}_{ij,\bom}(x_i,x_j') = A_{ij} e^{\bom x_i/v_W} + B_{ij}
e^{-\bom x_i/v_W}$ for $0<x_i<{\rm min}(x_j',L_{i1})~\de_{ij}+L_{i1}(1-
\de_{ij})$ ($v_W$ is the velocity in the wire region), then 
$B = - M ~A$, where $M$ is the current splitting matrix at the junction 
\cite{sandler,nayak,lal,chen,chamon,meden,das,giuliano,bellazzini1,agarwal}.

\noi The boundary condition in (iii) arises from the fact that the incoming 
and outgoing currents (and hence the bosonic fields) at the junction are 
related by the matrix $M$. Various constraints at the junction such as current
conservation and unitarity of the evolution of the system in real time (i.e., 
no power is dissipated exactly at the junction) imply that each row and column
of $M$ must add up to unity and that $M$ must be orthogonal. The possible $M$ 
matrices are restricted to two classes parameterized by a single parameter 
$\theta$~\cite{sandler,nayak,lal,chen,chamon,meden,das,giuliano,bellazzini1,
agarwal}: (a) det$(M_1) = 1$ and (b) det$(M_2) = -1$, which can be expressed as:
\bea M_1 = \left( \begin{array}{ccc} 
a & b & c \\
c & a & b \\
b & c & a \end{array} \right) &{\rm and}&
M_2 = \left( \begin{array}{ccc} 
b & a & c \\
a & c & b \\
c & b & a \end{array} \right),~~~ \label{m1-m2} \eea
where $a = (1+2\cos \theta)/3$ and $b(c) = (1-\cos \theta +(-) \sqrt{3} \sin 
\theta)/3$. We note that $(M_2)^2 = {\mbb 1}$ for any value of $\theta$; this 
relation will be used below.

Note that by introducing the orthogonal matrix $M$, we have made the 
simplifying assumption that there is no dissipation exactly at the junction.
It is calculationally simpler to separate the junction, which governs how the 
incoming currents are distributed amongst the different wires, from the 
regions of dissipation which lie away from the junction.

Solving eq.~\eqref{G-Eq-Mo} with the above boundary conditions and finally 
taking the limit $\bom \to 0 + \ep$, we get the following expression for the 
dc conductance matrix: 
\bea G &=& - ~\f{e^2 K_W}{\pi} ~[ {\mbb 1} + M + K_W ({\mbb 1} - M) ({\mbb 1} + \f{e^2}{\pi}
\mbb R)]^{-1} \nn \\
& & ~~~~~~~~\times ~[ {\mbb 1} - M], \label{cond} \eea
where $\mbb R$ is a $3 \times 3$ diagonal matrix with ${\mbb R}_{ii} = R_i =
r_{i0} (L_{i2}-L_{i1})$; note that $R_i$ is simply the total resistance in 
wire $i$. The conductance matrix relates the outgoing current $I_i$ to the 
potential $V_i$ applied in lead $i$ as $I_i = \sum_j G_{ij} V_j$. One can show
in general that each row and column of $G$ must add up to zero; the columns 
adding up to zero is a consequence of current conservation ($\sum_i I_i$ must 
be zero), while the rows must add up to zero because each of the $I_i$ must 
vanish if the $V_j$'s have the same values in all the wires.

One can show that the conductance of a single wire given in 
eq.~\ref{dc-cond} follows from eq.~\eqref{cond} if we choose $M = M_2$ with 
$\theta =0$, $2\pi/3$ or $4\pi/3$. For instance, if $\theta =0$, wire 3 
decouples from wires 1 and 2 (so that $G_{ij} = 0$ if either $i$ or $j =3$), 
while the conductance across wires 1 and 2 becomes independent of $K_W$ and 
is given by eq.~\eqref{dc-cond}, with $R=r_{10} (L_{12}-L_{11}) + r_{20} 
(L_{22}-L_{21})$ being the total resistance in wires 1 and 2. 

Eq.~\eqref{cond} can also be derived in general using the 
equation of motion approach in the $\om \to 0$ limit in the same way
as described above for the single wire case. We find that the precise
profiles of $K(x_i)$, $v(x_i)$ and $r(x_i)$ in the different wires 
are not important; all that matters is that the values of $K$ and $v$
are given by $K_W, ~v_W$ as $x_i \to 0+\ep$ and by $1, ~v_F$ as $x_i 
\to \infty$, and that the diagonal elements of ${\mbb R}$ are given
by $R_i = \int dx_i r (x_i)$.

{\it Conductance for the $M_1$ class.-}
In the $M_1$ class, the case $\theta =0$ is trivial because $M_1 (0) ={\mbb 1}$
and $G=0$. Let us now consider other values of $\theta$. We find that in
general $G$ depends on $K_W$, $\theta$, and the resistances $R_i =
r_{i0} (L_{i2}-L_{i1})$. [An exception arises for the case $\theta = \pi$
where we find that $G$ is independent of $K_W$ and depends only on the $R_i$.
This occurs whenever $M^2 = {\mbb 1}$ which is true for $M_1 (\pi)$ and also 
for the $M_2$ class for any $\theta$ as discussed below.]
The dependence of $G$ on $K_W$ for the $M_1$ class is to be contrasted to 
the case of a single wire where the conductance is independent of $K_W$
\cite{maslov,ponomarenko,safi1,safi3,thomale}. In fig.~\ref{diagram}
(b), we show the matrix elements $G_{11}$, $G_{12}$ and $G_{13}$ as functions
of $K_W$ for the case of $M_1$ with $\theta = 2\pi/3$ and ${\mbb R = 0}$.

In the limit that $R_i \to \infty$ (which is physically relevant when 
$R_i >> \pi/e^2$), we find that the conductance matrix takes the simple form
\bea \hspace*{-.7cm} G &=& \f{1}{R_1 R_2 + R_2 R_3 + R_3 R_1} \nn \\
\hspace*{-.7cm} & & \times \left(\begin{array}{ccc}
- R_2 - R_3 & R_3 & R_2 \\
R_3 & - R_1 - R_3 & R_1 \\
R_2 & R_1 & - R_1 - R_2 \end{array}\right), \label{r-inf-m1} \eea
which is independent of both $K_W$ and $\theta$. Interestingly, the form in
eq.~\eqref{r-inf-m1} is exactly the same as that obtained for a classical
system in which three wires with resistances $R_i$ meet at
a junction. If a potential $V_i$ is applied to wire $i$, and the potential
at the junction is $V_0$, then the outgoing currents are given by $I_i =
(V_0 - V_i)/R_i$ for all $i$. Using Kirchoff's circuit laws to eliminate $V_0$,
we find that the conductance matrix relating $I_i$ to $V_j$ is given by
eq.~\eqref{r-inf-m1}.
 
{\it Conductance for the $M_2$ class.-}
In this case the property $M^2 = {\mbb 1}$ combined with eq.~\eqref{cond} 
can be used to prove that $G$ is independent of $K_W$ for any choice of 
$\theta$ and $R_i$. The exact expression for $G$ turns out to be
\bea G &=& - ~\f{e^2}{\pi} ~\f{3 ({\mbb 1}- M_2)}{D}, \nn \\
{\rm where} ~~D &=& 2(\vr_1+\vr_2+\vr_3) + \cos \theta (\vr_1+\vr_2-2\vr_3)
\nn \\
& & -\sqrt{3}\sin \theta (\vr_1-\vr_2), \label{r-inf-m2}\eea
where $\vr_i = 1 + (e^2/\pi) R_i$. 
We can see that $G$ does not depend on $K_W$.

{\it Time reversal invariance.-}
It is interesting to look at our results from the point of view of time
reversal ($\cal T$) invariance. There are two sources of $\cal T$ 
breaking in our system:

\noi (i) the presence of resistances (i.e., dissipation) clearly violates 
$\cal T$. This is evident from eq.~\eqref{eom} which is not invariant under 
$t \to - t$.

\noi (ii) the current splitting matrix $M$ is $\cal T$ invariant only if it 
is symmetric. This is because the outgoing and incoming currents near the
junction satisfy $I_{out} = M I_{in}$, while $\cal T$ interchanges $I_{in}$
and $I_{out}$. We then see that $I_{in} = M I_{out}$ is satisfied only if 
$M^{-1} = M^T = M$. 

We observe that $M_2$ is symmetric and therefore
$\cal T$ invariant for all values of $\theta$, while $M_1$ is $\cal T$ 
invariant only if $\theta = 0$ or $\pi$. A junction described by $M_1$ 
can  exist only if $\cal T$ is broken, for instance, by applying a magnetic 
field through the junction, assuming that this has a finite cross-section.

The two sources of $\cal T$ breaking mentioned above are not related to 
each other since one acts at the resistances and the other acts only at the
junction. However, we showed above that if $M$ is symmetric, i.e., is $\cal 
T$ invariant, then the conductance $G$ is independent of $K_W$, regardless 
of the values of the resistances. Thus there is a remarkable connection 
between $\cal T$ breaking at the junction and the dependence of $G$ on
the interaction parameter $K_W$.

{\it Discussion.-}
To summarize, we have presented a phenomenological formalism which allows
us to study the effect of resistive regions in a quantum wire using the 
language of bosonization. This enables us to calculate the conductance of 
systems in which both the interaction parameter $K$ and the resistivity $r$
vary with $x$. The bosonic equation of motion makes it possible to visualize 
what happens when a charge density pulse is incident on the resistive and 
interacting regions. Finally, by introducing a current splitting matrix $M$ 
to describe a junction, we have extended the analysis to a three-wire system. 
We find that $G$ depends on $K_W$ for the class $M_1$ (except for the
special case with $\theta = \pi$), but not for the class $M_2$ in which
case we have found an analytical expression for $G$. Thus we have 
generalized the well-known results of Safi-Schulz and Maslov-Stone to 
include systems with junctions and resistances. It may be possible to test 
our results experimentally by, for instance, varying $K_W$ and $\theta$ by 
applying a gate voltage and a magnetic field near a junction of quantum wires
~\cite{chamon,das}, and measuring how this changes the conductance matrix.

\acknowledgments
We thank Sourin Das for stimulating discussions. A. S. thanks Abhishek R. Bhat
for help with the numerics. A. S. thanks CSIR, India for financial support, 
and D. S. thanks DST, India for financial support under Project No. 
SR/S2/CMP-27/2006.

\end{document}